\documentclass[a4paper,twocolumn]{article}
\usepackage[utf8]{inputenc}
\usepackage[T1]{fontenc}
\usepackage[numbers,compress,merge]{natbib}
\usepackage[colorlinks=true]{hyperref}
\usepackage{graphicx}
\usepackage{amsmath,amssymb}
\usepackage[varg]{txfonts}
\usepackage[affil-it]{authblk}

\title{Scalar absorption cross section of rotating black holes with tidal charge}

\author{Ednilton S. de Oliveira\footnote{ednilton@ufpa.br}}
\affil{Faculdade de F\'isica, Universidade Federal do Par\'a,
	66075-110, Bel\'em, Par\'a, Brazil}

\date{\today}

\begin{document}

\maketitle

\begin{abstract}
	
	The absorption cross section of rotating black holes in the Randall-Sundrum brane-world scenario for massless scalar waves is obtained via the partial-wave method. A variety of incidence directions with respect to the black hole rotation axis is considered. Special attention is given to near-extreme black holes, in which cases the absorption cross section increases as the tidal charge becomes more negative. It is shown that black holes with different configurations can present capture cross sections/shadows of same size. However, in such cases, the absorption spectra are different mainly for small frequencies. Numeric results are compared with the capture cross sections in the high-frequency limit, and with the black hole area in the low-frequency limit showing good agreement.
	
\end{abstract}

\maketitle

\section{Introduction}

General Relativity has proven once again to be a successful theory of gravity with the recent first observation of a black hole shadow~\cite{eht2019aj_l1,*eht2019aj_l2,*eht2019aj_l3,*eht2019aj_l4,*eht2019aj_l5,*eht2019aj_l6}, and the detections of gravitational waves~\cite{LIGO2019prx9_031040,LIGO2020ajl892_l3,*LIGO2020,*LIGO2020aj896_l44}. Despite this, some open issues, e.g. dark energy, dark matter, and the hierarchy problem, leave space for the rise of alternative models of gravity. A set of these alternatives initially addressed to solve the hierarchy problem includes the brane-world models which state gravitons and other unknown fields propagate in the bulk spacetime with more than four dimensions, while Standard-Model particles are confined to our four-dimensional usually observed universe, the brane~\cite{Arkani1998plb429_263,Antoniadis1998plb436_257,Randall1999prl83_3370}. One of the most interesting consequences of such models is the possible creation of microscopic black holes in particle collisions with center-of-mass energy higher than the reduced Planck scale, which can be of only a few TeVs~\cite{Park2012ppnp67_617,Mack2020jhep20_187}.

The possibility of black holes being created on particle collisions shows that these objects are not only important in the astrophysical level, but may be the key to identify the effects of extra dimensions in our four-dimensional world even in lab based experiments. In such panorama, part of the problem then reduces to describing black holes in brane-world models as well as studying their physical properties, specially their evaporation spectra via Hawking radiation~\cite{Hawking1975cmp43_199} which should provide the evidences that physicists have been seeking for. This has motivated the determination of the evaporation rate of extra-dimensional black holes on the brane~\cite{Emparan2000prl85_499,Kanti2003prd67_104019,Harris2003jhep10_014,Kanti2005prd71_104002,Ida2005prd71_124039,Ida2006prd73_124022,Harris2005plb633_106,Duffy2005jhep09_049,Casals2006jhep02_051,Casals2006jhep03_019}, as well as their greybody factors~\cite{Kanti2002prd66_024023,Ida2003prd67_064025,Creek2007prd75_084043,Kanti2014prd90_124077} and physical quantities related to them, as the absorption~\cite{Jung2004jhep09_005,Jung2005npb717_272} and scattering~\cite{deOliveira2017cqg35_065007,deOliveira2018epjc78_876,deOliveira2019} cross sections.

One of the solutions found in the Randall-Sundrum scenario~\cite{Randall1999prl83_3370,Randall1999prl83_4690} can be put in the form of the Reissner-Nordström metric, but with $Q^2$, which is associated with the squared electric charge in the Reissner-Nordström solution~\cite{Chandra1983}, describing now a manifestation of the fifth-dimension influence on the brane~\cite{Dadhich2000plb487_1}. However, in the context of the Randall-Sundrum models, $Q^2$ can assume negative values and, as argued in Ref.~\cite{Dadhich2000plb487_1}, this actually represents the most natural case. In conformity, recent researches based on the analysis of the shadow of the galaxy M87 core have concluded that if a positive tidal charge exists, it should be very small~\cite{Banerjee2020prd101_041301,Neves2020doc}. The analysis of that shadow has also been used to constrain the curvature radius of the extra dimension~\cite{Vagnozzi2019prd100_024020}.

The negative tidal charge strengths the gravitational interaction so that static black holes endowed with it have a bigger event horizon when compared with Schwarzschild black holes. As consequences, they also present bigger absorption cross sections~\cite{Toshmatov2016prd93_124017,deOliveira2018epjc78_876,deOliveira2019}, photon spheres~\cite{Abdujabbarov2017prd96_084017}, and shadows~\cite{Amarilla2011prd85_064019}. It has also been shown that such black holes are stable against bosonic oscillations~\cite{Molina2016prd93_124068,Toshmatov2016prd93_124017}, which present smaller frequencies and higher damping rates for black holes with more negative tidal charges~\cite{Toshmatov2016prd93_124017}.

Similarly to what happens in the static case, the metric of (electrically neutral) rotating black holes with tidal charge has the same form of electrically charged rotating black holes of General Relativity with $Q^2$ possibly assuming negative values~\cite{Aliev2005prd71_104027}. In such cases, one of the most interesting consequences is that the black holes are allowed to spin faster than extreme Kerr black holes. We have shown in Ref.~\cite{deOliveira2020lzp} that this has important effects on superradiance, making near-extreme~\footnote{Here we focus mainly on near-extreme black holes as their formation from the collapse of astrophysical objetcs is believed to be favoured~\cite{Misner_etal1973}.} rotating black holes with negative tidal charges superradiate more than Kerr black holes. A similar conclusion can be obtained considering the Penrose process~\cite{Khan2019pdu26_100331}. Superradiance of massless scalar waves can also be crucially modified around rotating black holes with negative tidal charges once maximum amplification can be retrieved from $m > 1$, where $m$ is the azimuthal quantum number, while in the case of Kerr black holes the same scattering presents maximum amplification for $m = 1$~\cite{Brito2015lnp906}.

Since their appearance, rotating black holes with tidal charge have been subject of many studies. Apart from the previously cited works in this context, which evolve superradiance, the Penrose process, and the constraining of the tidal charge via the shadow of the galaxy M87 core, other works concerning this system address to the determination of ISCOs~\cite{Aliev2005prd71_104027,Abdujabbarov2010prd81_044022,Pun2008prd78_084015}, shadows~\cite{Schee2008ijmpd18_983,Amarilla2011prd85_064019,Abdujabbarov2017prd96_084017,Eiroa2017epjc78_91}, light deflections and frame dragging~\cite{Aliev2009prd80_044023}, and orbital resonances~\cite{Stuchlik2009grg41_1305}.

In the present work we address to the problem of the absorption cross section of massless scalar waves impinging upon rotating black holes with negative tidal charges from different directions relative to the axis of rotation. In order to do so, we apply the partial-wave method and solve the massless Klein-Gordon equation numerically. We also make use of the geodesic analysis to determine the capture cross sections and shadows cast by the black hole, and show that even though black holes with different configurations can present the same shadow, their absorption cross sections differ. The capture cross section together with the low-frequency absorption cross section of axially symmetric black holes determined by Higuchi~\cite{Higuchi2001cqg18_L139} help us to verify the consistency of our numerical results.

This work is based on the follow structure: In Sec~\ref{sec:sys} we describe the propagation of the massless scalar waves around rotating black holes with tidal charge. The forms of the absorption cross sections, including the one obtained from the partial-wave method, low- and high-frequency analysis are presented in Sec.~\ref{sec:abs}. Our main results obtained via the partial-wave method are presented in Sec.~\ref{sec:results}, as well as their comparisons with the approximations described in the previous section. In Sec.~\ref{sec:conc} we conclude with our final remarks. Here we use $c = G = \hbar = 1 $.

\section{Wave-propagation description}
\label{sec:sys}

\subsection{The black hole}

Rotating black holes with tidal charge appeared first in Ref.~\cite{Aliev2005prd71_104027}. There we can find a complete analysis of their spacetime structure. Here we restrict ourselves to review succinctly the main properties of these systems. In Boyer-Lindquist coordinates, such black holes are described by:
\begin{eqnarray}
	ds^2 & = & \left(1-\frac{2Mr-b}{\rho^2}\right) dt^2 + 2a\frac{2Mr-b}{\rho^2} \sin^2\theta \, dt d\phi \nonumber \\ &&  - \frac{\rho^2}{\Delta} dr^2 - \rho^2 d\theta^2\nonumber \\
	&& - \left(r^2 + a^2 + \frac{2Mr-b}{\rho^2} a^2 \sin^2\theta \right) \sin^2 \theta\, d\phi^2,
	\label{ds}
\end{eqnarray}
where
$$ \Delta = r^2 -2Mr + a^2 + b,$$
$$ \rho^2 = r^2 + a^2\cos^2\theta,$$
$M$ is the black holes mass, $a$ is its angular momentum per unit mass, and $b \equiv qM^2$ is the tidal charge. If $q > 0$, then the metric above coincides with the Kerr-Newman metric~\cite{Newman1965jmp6_918}, but in the Randall-Sundrum brane-world scenario $q$ is expected to be negative~\cite{Dadhich2000plb487_1}.

The event horizon of a rotating black hole with tidal charge is located at:
\begin{equation}
	r_+ = M + \sqrt{M^2-a^2 - b},
	\label{horizons}
\end{equation}
and the ergoregion at:
\begin{equation}
	r_\text{ergo} (\theta)= M + \sqrt{M^2-a^2\cos^2\theta -b}.
	\label{ergo}
\end{equation}
The maximum angular momentum per unit mass that rotating black holes with tidal charge can acquire is determined by
\begin{equation}
	a_c = M\sqrt{1 - q}.
	\label{ac}
\end{equation}
Since $q <0$, this means that these black holes can spin faster than Kerr black holes of same mass. As a consequence, a negative tidal charge can make a rotating black hole to get more energetic in which concerns to the Penrose process~\cite{Khan2019pdu26_100331} and superradiance~\cite{deOliveira2020lzp}. The angular velocity of the event horizon is:
\begin{equation}
	\Omega_H \equiv \frac{a}{r_+^2 + a^2}.
	\label{Omega}
\end{equation}

\subsection{Massless scalar waves}

Here we consider a monochromatic massless scalar plane wave being scattered by the black holes described by Eq.~\eqref{ds}. Such waves are governed by the equation:
\begin{equation}
	\frac{1}{\sqrt{-g}} \partial_{\mu} \left ( \sqrt{-g} g^{\mu\nu} \partial_\nu \Phi \right)  = 0.
	\label{KG}
\end{equation}

The separability of this equation can be done if
$$
\Phi = \frac{R_{\omega lm}(r)}{\sqrt{r^2+a^2}} S_{\omega lm}(\theta) e^{i(m\phi-\omega t)}.
$$
From this we obtain the following radial equation:
\begin{equation}
	\frac{d^2}{dr_*^2}R_{\omega lm} + \left[\omega^2 - V_{\omega lm}(r)\right]R_{\omega lm} = 0,
	\label{radial_eq}
\end{equation}
where
$$
\frac{d}{dr_*} = \frac{\Delta}{r^2+a^2} \frac{d}{dr}$$
defines the tortoise coordinate $r_*$ and
\begin{eqnarray}
	V_{\omega lm} & = & -\frac{1}{(r^2+a^2)^2} \left[m^2a^2-\Delta(\lambda_{lm}+\omega^2a^2) \right. \nonumber\\
	&& \left. + 2ma\omega(b - 2Mr)\right] + \Delta\frac{\Delta+2r(r-M)}{(r^2+a^2)^3} \nonumber\\
	&& -\frac{3r^2\Delta^2}{(r^2+a^2)^4}.
	\label{V}
\end{eqnarray}
Here, $\lambda_{lm}$ are the eigenvalues of the oblate spheroidal harmonics $S_{\omega lm}$~\cite{Abramowitz_etal1964} which obey the equation:
\begin{equation}
	\frac{d}{dx} \left((1-x^2) \frac{dS_{\omega lm}}{dx} \right) + \left[(a\omega)^2 x^2 - \frac{m^2}
	{1-x^2} + \lambda_{lm}\right] S_{\omega lm} = 0,
	\label{S}
\end{equation}
with $x = \cos\theta$. We adopt the normalization of the spheroidal harmonics as~\cite{Macedo2013prd88_064033}:
\begin{equation}
	2\pi \int |S_{\omega lm}(\theta)|^2 \sin\theta\,  d\theta = 1.
	\label{Snorm}
\end{equation}

The radial equation possesses approximated forms near the event horizon and far from the black hole. For $r \gtrsim r_+$, we have:
\begin{equation}
	R_{\omega lm} \sim A_\text{tr} \exp\left[-i(\omega - m\Omega_H) r_*\right], 
	\label{Rhor}
\end{equation}
and for $r \gg r_+$,
\begin{eqnarray}
	R_{\omega l m} & \approx & \sqrt{\frac{\pi\omega r_*}{2}}\left[(-i)^{\nu+1/2} A_\text{inc} H_{\nu}^{(2)} (\omega r_*) + \right. \nonumber \\
	&& \left. i^{\nu+1/2} A_\text{ref} H_\nu^{(1)} (\omega r_*) \right],
	\label{asy_hankel}
\end{eqnarray}
where $H_\nu^{(1),(2)}$ are the Hankel functions~\cite{Abramowitz_etal1964} with $\nu = \sqrt{a^2\omega^2 + \lambda_{lm} + 1/4}$. Here, $A_\text{tr}, A_\text{inc}, A_\text{ref}$ are the transmitted, incident, and reflected amplitudes, respectively. Considering the asymptotic forms of the Hankel functions, we can show that for $r \to \infty$:
\begin{equation}
	R_{\omega lm} \sim A_\text{inc} e^{-i \omega r_*} +  A_\text{ref} e^{i \omega r_*}.
	\label{Rinf}
\end{equation}

Greybody factors determine the amount of absorbed radiation. They are define as:
\begin{equation}
	\gamma_{lm} = \left(1 - \frac{m\Omega_H}{\omega} \right) \frac{|A_\text{tr}|^2}{|A_\text{inc}|^2} = 1- \frac{|A_\text{ref}|^2}{|A_\text{inc}|^2}.
	\label{gbf}
\end{equation}
As a remark, in the cases $\omega < m\Omega_H$, the reflection coefficient $|A_\text{ref}/A_\text{inc}|^2$ is higher than unity as a result of the fact that such partial wave carries out more energy than it had before the scattering; this is superradiance~\cite{Brito2015lnp906}. 

\section{Massless absorption}
\label{sec:abs}

The total absorption cross section for a monochromatic massless scalar plane wave initially propagating at the direction $\gamma$ with relation to the black hole axis of rotation is given by~\cite{Dolan_2006-thesis}:
\begin{equation}
	\sigma = \sum\limits_{l=0}^{\infty} {\sum\limits_{m = -l}^{l} \sigma_{lm}},
	\label{tacs}
\end{equation}
where $\sigma_{lm}$ are the partial absorption cross sections defined as:
\begin{equation}
	\sigma_{lm} = \frac{4\pi^2}{\omega^2} |S_{\omega l m }(\gamma) |^2 \gamma_{lm}.
	\label{pacs}
\end{equation}

The determination of the absorption cross section of black holes with tidal charge in the full-frequency domain requires numerical evaluation of the radial equation~\eqref{radial_eq} and the oblate spheroidal harmonics. However, analytical approximations are useful to verify the consistency of the numeric approach, but are restricted to low- or high-frequency regimes. We briefly describe these approximations below.

\subsection{Low-frequency analysis}

We have shown in Ref.~\cite{deOliveira2020lzp} that it is possible to obtain a low-frequency approximation for the greybody factors if we consider a low-frequency solution for the radial equation~\eqref{radial_eq}. Unfortunately, such solution is only valid for $ma \neq 0$ and consequently it cannot be used to describe the low-frequency approximation of the absorption cross section, despite the fact it can be successfully applied to obtain approximated amplification factors describing superradiant scattering. This happens because the main contribution to the low-frequency absorption cross section for massless scalar plane waves comes from the $l = 0$ mode.

Nevertheless, it is possible to circumvent this difficulty once stationary black holes obey the universality of the absorption cross section of massless scalar waves obtained by Higuchi~\cite{Higuchi2001cqg18_L139}. This universality states that the low-frequency absorption cross sections in such cases equal the black hole area. Therefore, in the limit $\omega \to 0$,
\begin{equation}
	\sigma \to \mathcal{A},
	\label{lf_abs}
\end{equation}
where $\mathcal{A}$ is the black hole area:
\begin{equation}
	\mathcal{A} \equiv 4\pi(2Mr_+-b).
	\label{A}
\end{equation}

\subsection{High-frequency analysis}

At very high frequencies, the absorption cross section tends to its classical limit, which is the capture cross section of geodesics. The area of the capture cross section coincides with the area of the shadow as measured in celestial coordinates by an observer at infinity.

The key equations to determine the shadows of rotating black holes described by~\eqref{ds} are~\cite{Aliev2009prd80_044023}:
\begin{equation}
	\rho^2 \dot{r} = \sqrt{\mathcal{R}},
	\label{dotr}
\end{equation}
and
\begin{equation}
	\rho^2 \dot{\theta} = \sqrt{\Theta},
\end{equation}
where the dot expresses differentiation with respect to an affine parameter,
\begin{equation}
	\mathcal{R} = [(r^2+a^2)E -a L_z]^2 - \Delta [\mathcal{K} + (L_z - aE)^2],
	\label{R}
\end{equation}
and
\begin{equation}
	\Theta = \mathcal{K} + (a^2E^2-L_z^2\csc^2\theta)\cos^2\theta,
	\label{Th}
\end{equation}
with $E, L_z$ being constants of motion and $\mathcal{K}$ a separation constant. These constants are related to the celestial coordinates $\alpha$ and $\beta$ as~\cite{Amarilla2011prd85_064019}:
\begin{equation}
	\alpha = - \xi\csc\gamma ,
	\label{alpha}
\end{equation}
and
\begin{equation}
	\beta = \pm (\eta +a^2\cos^2\gamma - \xi^2\cot^2\gamma)^{1/2},
\end{equation}
where $\xi \equiv L_z/E$, $\eta \equiv \mathcal{K}/E^2$, and $\gamma$ defines the angular position of the observer at infinity.
% is the initial direction of the geodesic at infinity and $\chi$ is in the plane perpendicular to the incidence direction starting from the rotation axis in the co-rotating sense~\cite{Macedo2013prd88_064033}. The capture cross section is delimited by the critical impact parameter $b_c(\gamma, \chi)$ and it is given by:
% \begin{equation}
% \sigma_\text{geo}(\gamma) = \frac{1}{2} \int\limits_{-\pi}^{\pi} b_c^2(\gamma,\chi) d\chi.
% \label{ccs}
% \end{equation}

The shadow cast by the black holes is delimited by the critical rays, which are the frontier between captured and scattered rays. These critical rays are described by $\eta_\text{crit}(\xi)$ determined by the system~\cite{Zakharov2005nb}:
\begin{equation}
	\mathcal{R} = 0,
	\frac{\partial \mathcal{R}}{\partial r} = 0.
	\label{dR0}
\end{equation}
The condition $\Theta \ge 0$ defines the adequate interval of $\xi$ for the curve $\eta_\text{crit}(\xi)$ which represents the critical geodesics suitable to deline the shadow depending on the value of $\gamma$. 

\begin{figure}[!htb]
	\centering
	\includegraphics[width=0.48\textwidth]{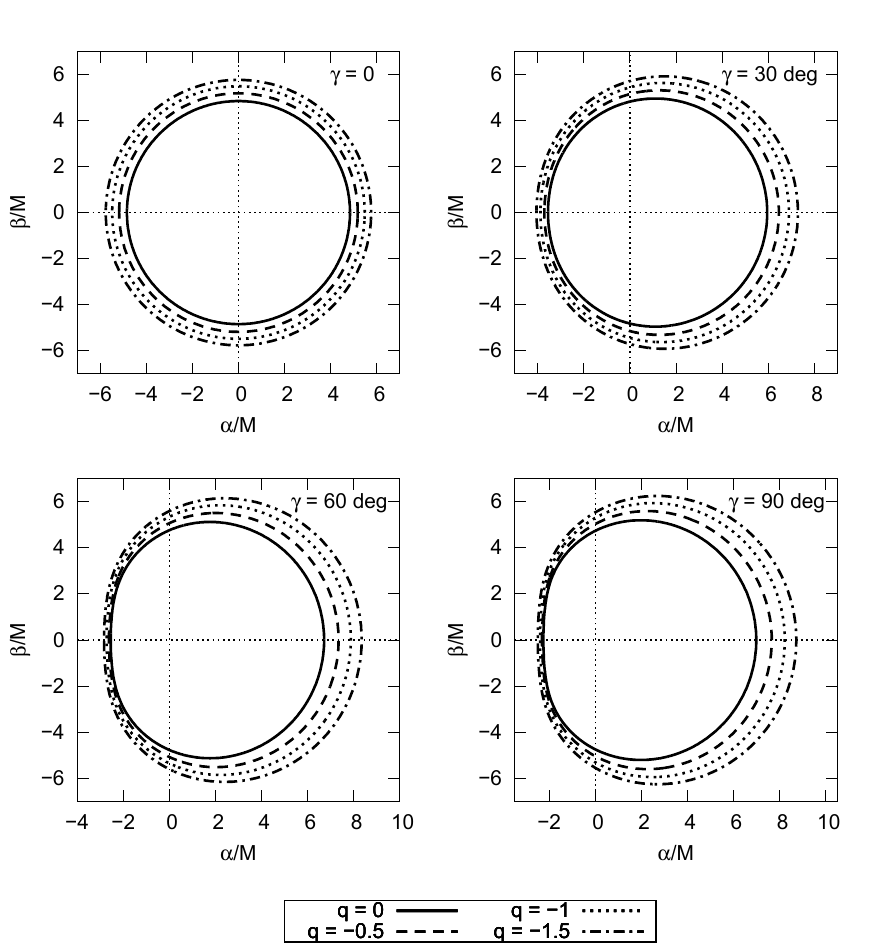}
	\caption{Shadows cast by near-extreme rotating black holes ($a = 0.99\, a_c$) with tidal charges $q = 0, -0.5, -1.0, -1.5$ as seen by an observer at infinity and $\gamma = 0$ (top-left), $\gamma = 30$ deg (top-right), $\gamma = 60$ deg (bottom-left), and $\gamma = 90$ deg (bottom-right).}
	\label{fig:shad_comp}
\end{figure}

Shadows cast by near-extreme rotating black holes with $a = 0.99 \, a_c$ and tidal charges $q = 0,-0.5,-1.0,-1.5$ are presented in Fig.~\ref{fig:shad_comp}. We see that the effect of the negative tidal charge is to increase the size of the shadow, while its shape does not seem to change significantly if the black hole rotation is kept near its extreme value (see Refs.~\cite{Neves2020doc,Amarilla2011prd85_064019,Schee2008ijmpd18_983} for shadows of rotating black holes with tidal charges considering further configurations).

\begin{figure}[!htb]
	\centering
	\includegraphics[width=0.48\textwidth]{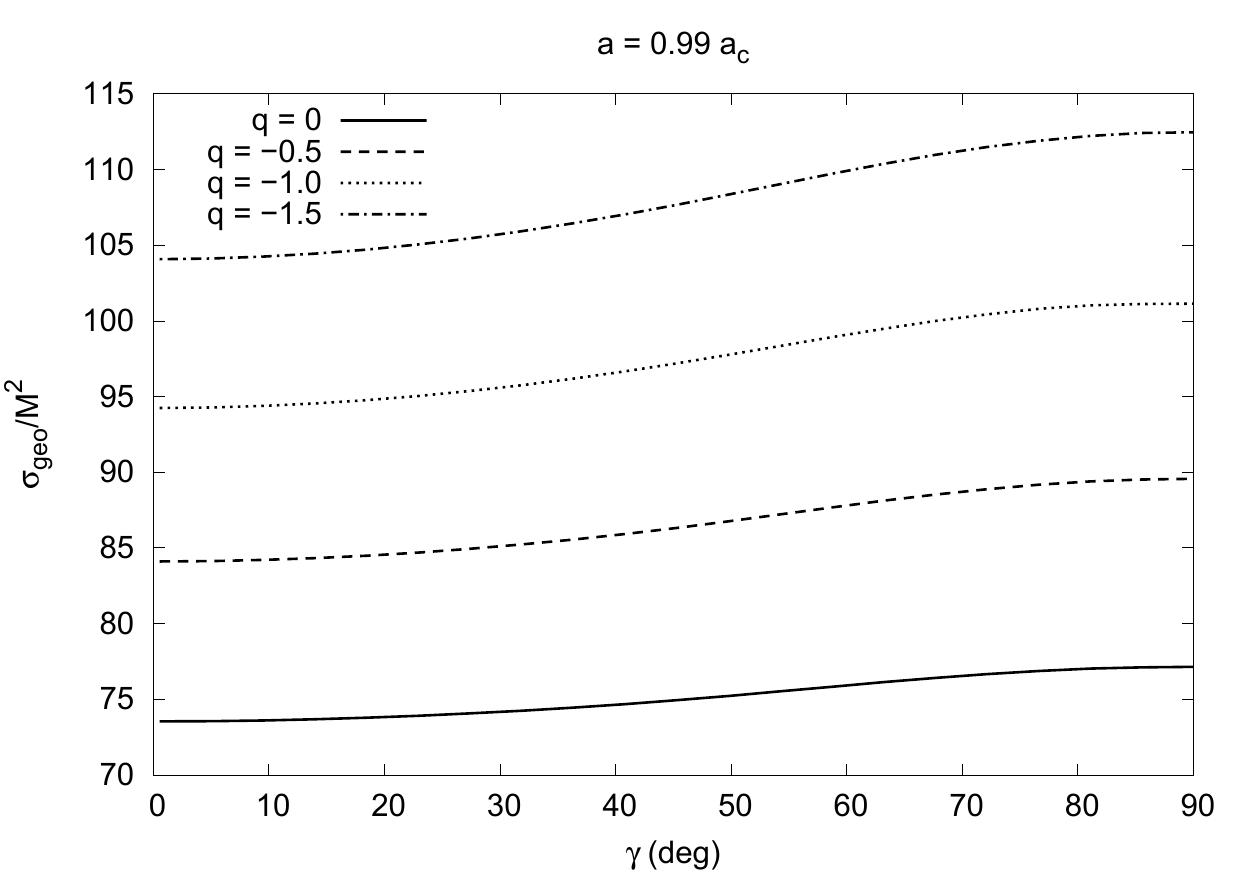}
	\caption{Capture cross section of near-extreme rotating black holes with tidal charge for massless particles varying with the angle of incidence. We consider the cases $q = 0, -0.5, -1.0, -1.5$.}
	\label{fig:ccs_comp}
\end{figure}

The capture cross sections, $\sigma_\text{geo}$, of near-extreme black holes with tidal charges $q = 0, -0.5, -1.0, -1.5$ varying with $\gamma$, which in this case plays the role of incidence direction, are presented in Fig.~\ref{fig:ccs_comp}. It becomes clearer that the increase of $\gamma$ is not only followed by a change at the form of the capture cross section/shadow, but also by an increase of its area.

\section{Results}
\label{sec:results}

Our main results are obtained numerically. This resumes to obtaining the oblate spheroidal harmonics and their eigenvalues numerically~\cite{Berti2006prd73_024013}. We also solve the radial equation~\eqref{radial_eq} numerically with boundary condition given by~\eqref{Rhor}, and then mach the solutions with the asymptotic forms in terms of the Hankel functions~\eqref{asy_hankel}. The sum of the total absorption cross section~\eqref{tacs} can be truncated at $l_\text{max}$, which depends on the desired frequency interval, $q,a$, and $\gamma$. Here, for example, for maximum frequency $M \omega_\text{max} = 1.5$, $\gamma = 90$ deg, $a = 0.99\, a_c$, $q = 0,-0.5,-1.0,-1.5$, we have chosen $l_\text{max} = 10,11,12,13$ respectively. This choice guarantees that the contributions of the neglected partial absorption cross sections are less than $0.1\%$ to the total absorption cross section at $M\omega_\text{max}$, and even smaller for $\omega < \omega_\text{max}$. If smaller values of $q,a,\gamma$ are chosen, than $l_\text{max}$ may be reduced.

\begin{figure}[!htb]
	\centering
	\includegraphics[width=0.48\textwidth]{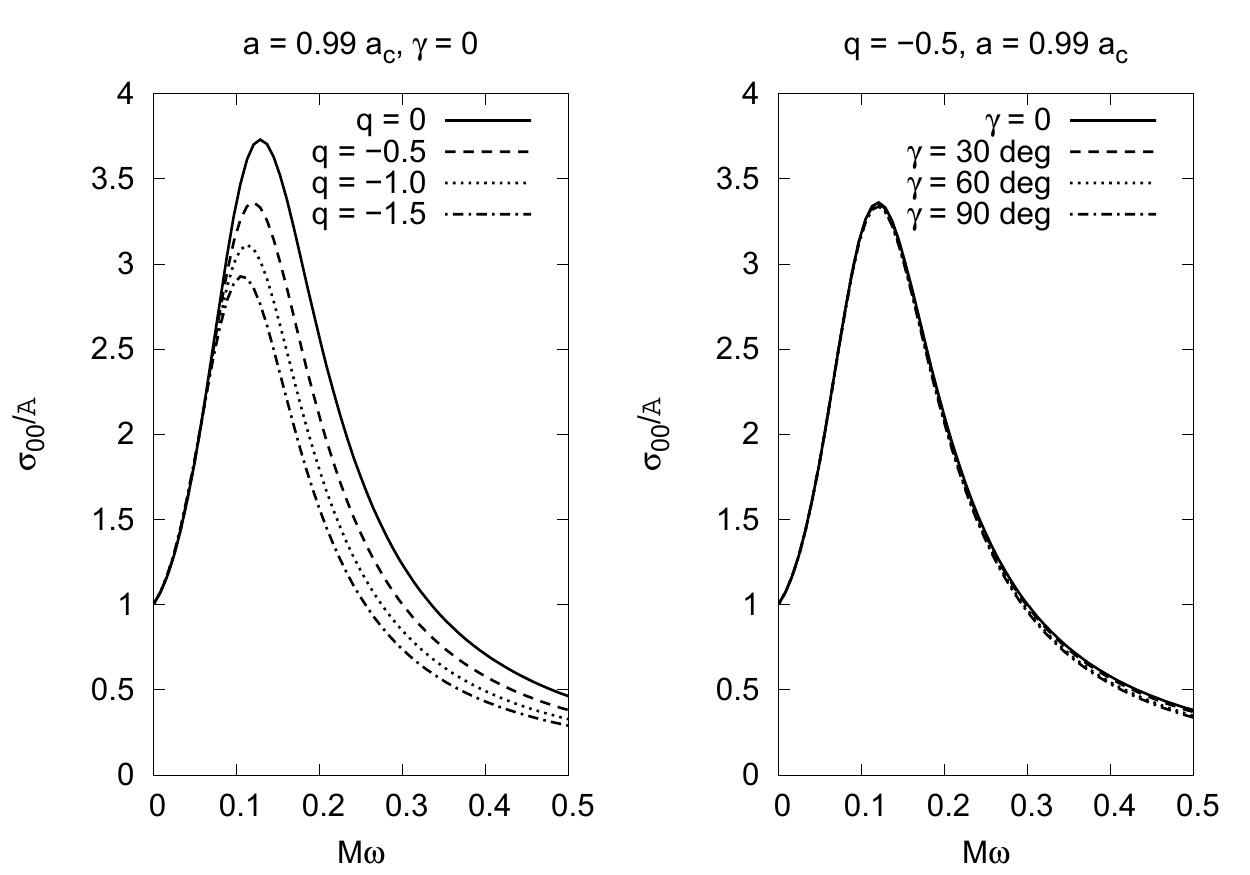}
	\caption{Partial scalar absorption cross section for $l = 0$ in units of the black hole area. \emph{Left:} the cases $q = 0, -0.5, -1.0, -1.5$ with $\gamma = 0$. \emph{Right:} the case $q = -0.5$ for directions of incidence $\gamma = 0$, 30 deg, 60 deg, 90 deg.}
	\label{fig:pacs0}
\end{figure}

Let us start by showing the partial absorption cross section of near-extreme rotating black holes with tidal charge in units of their area for the mode $l = 0$ in Fig.~\ref{fig:pacs0}. The results are compared for the cases $q = 0, -0.5, -1.0, -1.5$ with axial incidence, and for $q = -0.5$ with incidences along $\gamma = 0$, 30 deg, 60 deg, 90 deg. This shows that rotating black holes with tidal charge obey the universality of the low-frequency absorption cross section for massless scalar waves once $\sigma/\mathcal{A} \to 1$ in the regime $M\omega \to 0$ independently of the incidence direction. Therefore, our results are consistent with the low-frequency approximation~\cite{Higuchi2001cqg18_L139}. Right panel of Fig.~\ref{fig:pacs0} also shows that the absorption of the partial wave $l = 0$ is only slightly affected by changes at the plane wave incidence direction.

\begin{figure*}[!htb]
	\centering
	\includegraphics[width=\textwidth]{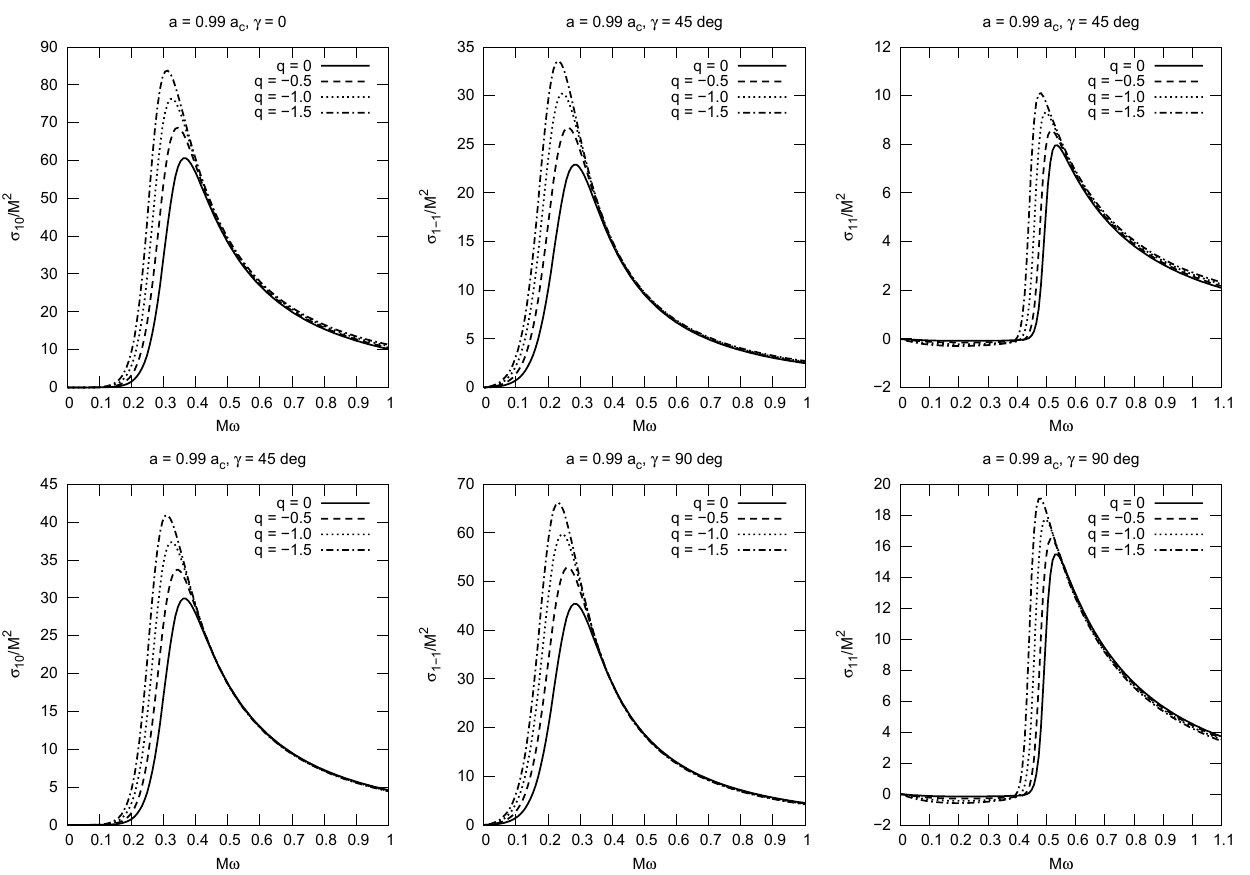}
	\caption{Partial scalar absorption cross section of near-extreme black holes with $q = 0, -0.5,-1.0, -1.5$ for $l = 1$. \emph{Left:} $m = 0$ for $\gamma = 0$ (top), $\gamma = 45$ deg (bottom); \emph{Center:} $m = -1$ for $\gamma = 45$ deg (top), $\gamma = 90$ deg (bottom); \emph{Right:} $m = 1$ for $\gamma = 45$ deg (top), $\gamma = 90$ deg (bottom).}
	\label{fig:pacs1}
\end{figure*}

In Fig.~\ref{fig:pacs1} we present the partial absorption cross sections for the cases $l = 1$, $m = -1,0,1$ considering near-extreme black holes with $q = 0, -0.5, -1.0, -1.5$. For the case $m = 0$ we have chosen incidences along $\gamma = 0$, 45 deg, and for $m = \pm 1$ we have chosen $\gamma = 45$ deg, 90 deg\footnote{Partial absorption cross sections for odd values of $l + m$ vanish if $\gamma = 90$ deg, while for $\gamma = 0$ the only non-vanishing contributions to the absorption cross sections come from $m = 0$.}. The presence of the negative tidal charge contributes to the enhancement of the partial absorption independently of the values of $l, m$, as well as the value of $\gamma$. However, the change at the incidence direction has different effects on different modes. Particularly in the cases presented in Fig.~\ref{fig:pacs1}, with the increase of $\gamma$, $\sigma_{10}$ decreases while $\sigma_{1\pm1}$ increase. We also call attention for the fact that co-rotating modes ($m > 0$) are less absorbed than their counter-rotating correspondents ($m < 0$).

\begin{figure}[!htb]
	\centering
	\includegraphics[width=0.49\textwidth]{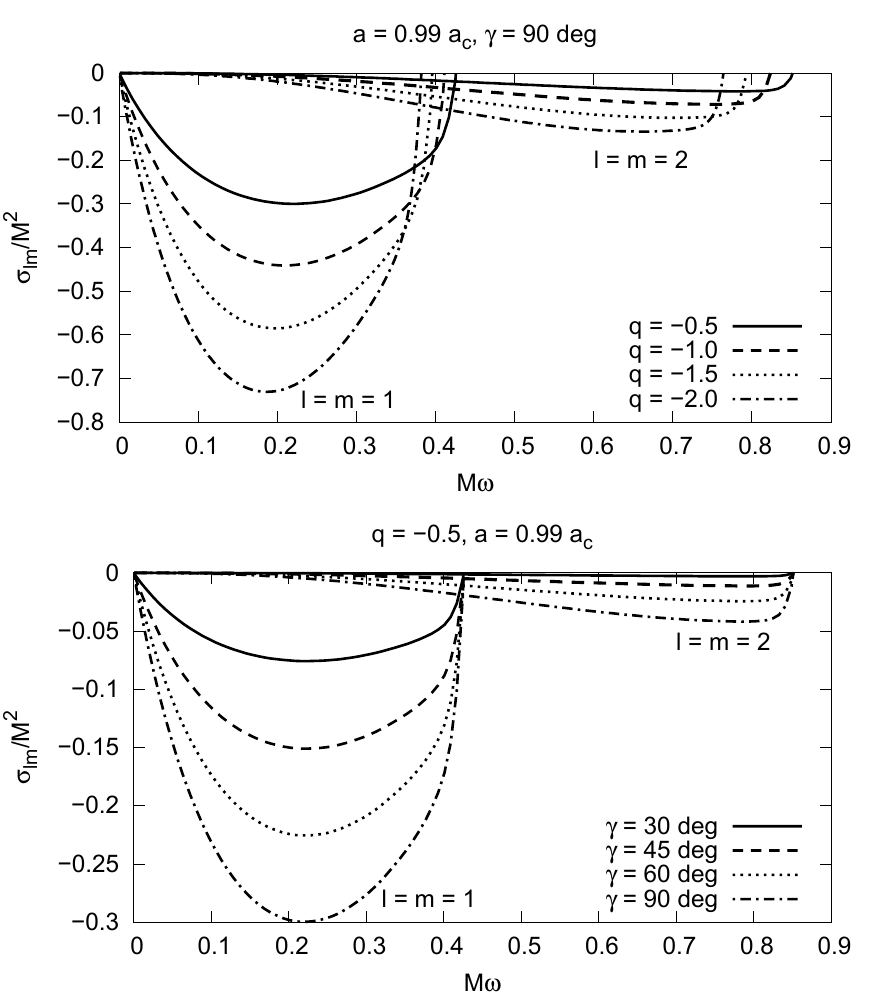}
	\caption{Zoom in the superradiant regime of the partial absorption cross sections for the cases $l = m = 1,2$. On the top panel we compare the cases in which $\gamma = 90$ deg for near-extreme black holes with tidal charges $q = -0.5, -1.0, -1.5, -2.0$. On the bottom panel we present the case with tidal charge $q = -0.5$ but for different incidence directions, namely $\gamma = 30$ deg, 45 deg, 60 deg, 90 deg.}
	\label{fig:sup_zoom}
\end{figure}

We see in Fig.~\ref{fig:pacs1} that the partial absorption cross section for $l = m = 1$ presents negative values for specific intervals of $M\omega$. This happens because of superradiance, as expected from Eqs.~\eqref{gbf} and~\eqref{pacs}, once $|A_\text{ref}/A_\text{inc}|^2 > 1$ in such cases. Figure~\ref{fig:sup_zoom} shows a zoom in the superradiant regime of the partial absorption cross section for the cases $l = m = 1,2$ considering near-extreme black holes with $q = -0.5, -1.0, -1.5, -2.0$ for incidence along $\gamma = 90$ deg, and with $q = -0.5$ for incidences along $\gamma = 30$ deg, 45 deg, 60 deg, 90 deg. The increase of the partial absorption cross sections due to the increase of $-q$ observed in Fig.~\ref{fig:pacs1} also contributes to the negative absorption enhancement in the superradiant regime. The negative partial absorption cross section also enhances when the angle of incidence increases towards 90 deg.

In Ref.~\cite{deOliveira2020lzp} we have shown that maximum amplification factor for black holes with $q = -2.0$ happens for $l = m =2$ instead of $l = m =1 $, which is the case of Kerr black holes, considering the scattering of massless scalar waves. Figure~\ref{fig:sup_zoom} shows that this cannot be inferred from the partial absorption cross section once maximum absorption in the superradiant regime gets highly suppressed as we go from $l = m  = 1$ to $l = m = 2$. This is due to the presence of the term $1/\omega^2$ in the partial absorption cross section~\eqref{pacs} combined with the fact that maximum amplification of each mode happens for higher frequencies as higher is the value of $m$.

\begin{figure}[!htb]
	\centering
	\includegraphics[width=0.48\textwidth]{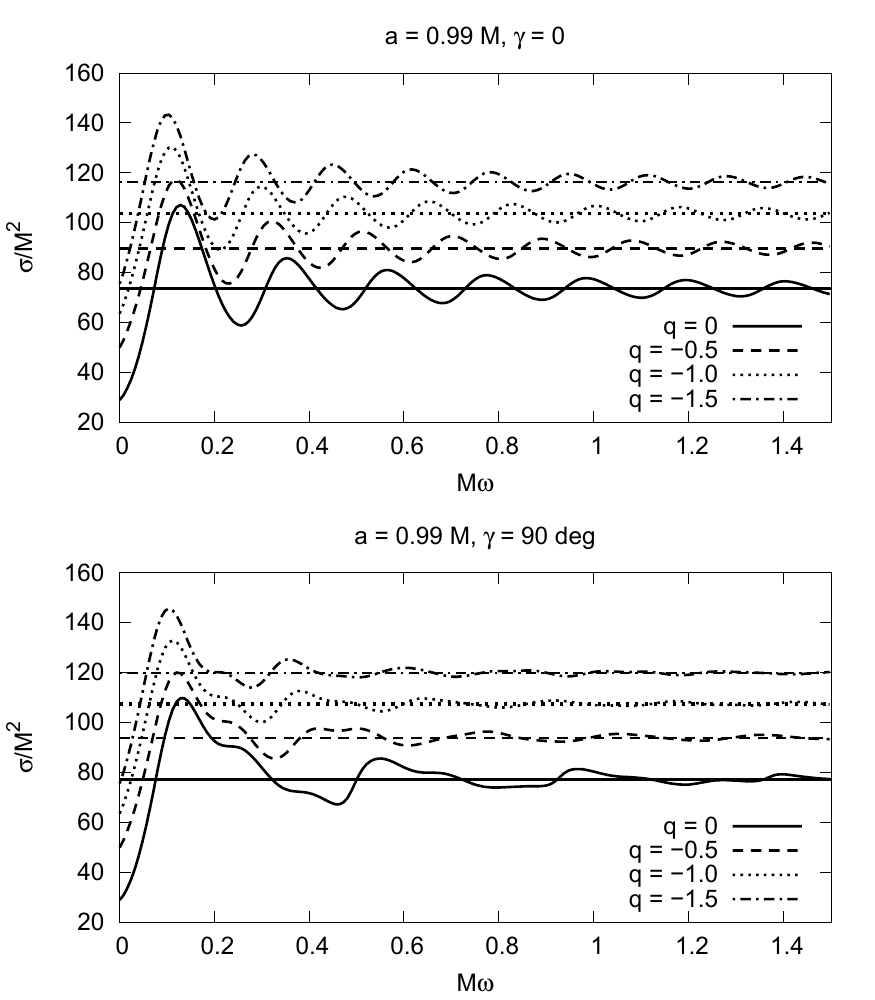}
	\caption{Total absorption cross section for massless scalar waves scattered by rotating black holes with $a = 0.99 M$ and tidal charges $q= 0, -0.5, -1.0, -1.5$. The angles of incidence are $\gamma = 0$ (top) and $\gamma = 90$ deg (bottom). The straight lines represent the respective values of the capture cross sections for null geodesics.}
	\label{fig:a0.99acs}
\end{figure}

Figure~\ref{fig:a0.99acs} shows the total absorption cross sections of rotating black holes with $a = 0.99M$ and $q = 0, -0.5, -1.0, -1.5$ for $\gamma = 0, 90$ deg compared with the respective capture cross sections (straight lines). In all cases, the absorption cross section tends to oscillate around its high-frequency corresponding value as $M\omega$ increases. This shows that our numeric results are consistent with the geodesic analysis of the capture cross sections/shadows. The variations of the absorption cross sections around their corresponding capture cross section values tend to be less regular as $\gamma$ goes from 0 to 90 deg. Such irregularity is partially due to the fact that co- and counter-rotating modes are absorbed differently, as observed in the study of the scalar absorption by Kerr black holes in Ref.~\cite{Macedo2013prd88_064033}. Despite this lack of regularity, the absorption cross section for the case $\gamma  = 90$ deg seems to converge more rapidly to the value of the corresponding capture cross section.

\begin{figure*}[!htb]
	\centering
	\includegraphics[width=\textwidth]{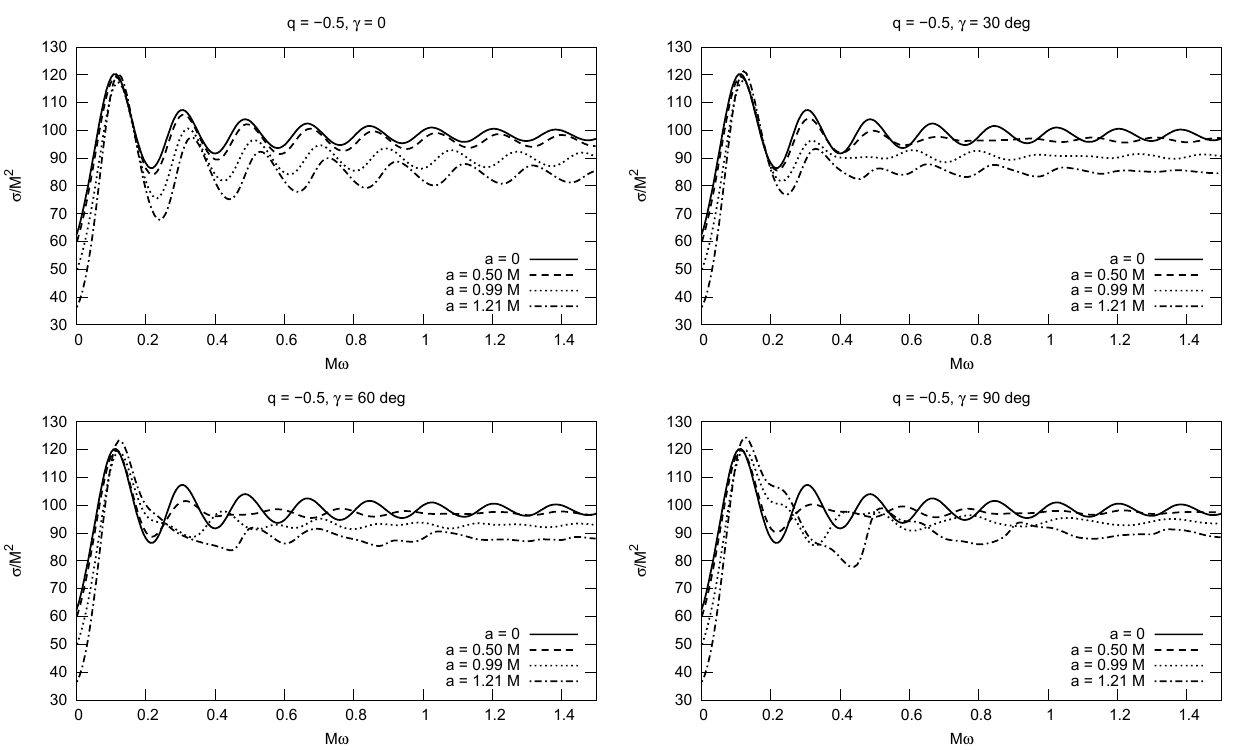}
	\caption{Total scalar absorption cross section of black holes with tidal charge $q = -0.5$, and $a = 0, 0.5 M, 0.99 M$, and $a = 1.21 M$ ($a \approx 0.99\,a_c$) for incidence directions $\gamma = 0$ (top-left), $\gamma = 30$ deg (top-right), $\gamma = 60$ deg (bottom-left), $\gamma = 90$ deg (bottom-right). }
	\label{fig:a_acs}
\end{figure*}

Comparisons of the scalar absorption cross sections of black holes with tidal charge $q = -0.5$ and different angular momenta, including the static case which has been studied numerically in Ref.~\cite{deOliveira2018epjc78_876}, for incidence directions $\gamma = 0$, 30 deg, 60 deg, 90 deg is presented in Fig.~\ref{fig:a_acs}. We see that rotation does not affect the absorption spectrum profile for axial incidence ($\gamma = 0$), but its increase contributes to the decrease of the absorption in average. When we consider off-axis incidence, both average and profile of the absorption cross section are modified by the increase of $a$, remarking the fact that absorption does not vary with $\gamma$ in the static case.

\begin{figure*}[!htb]
	\centering
	\includegraphics[width=\textwidth]{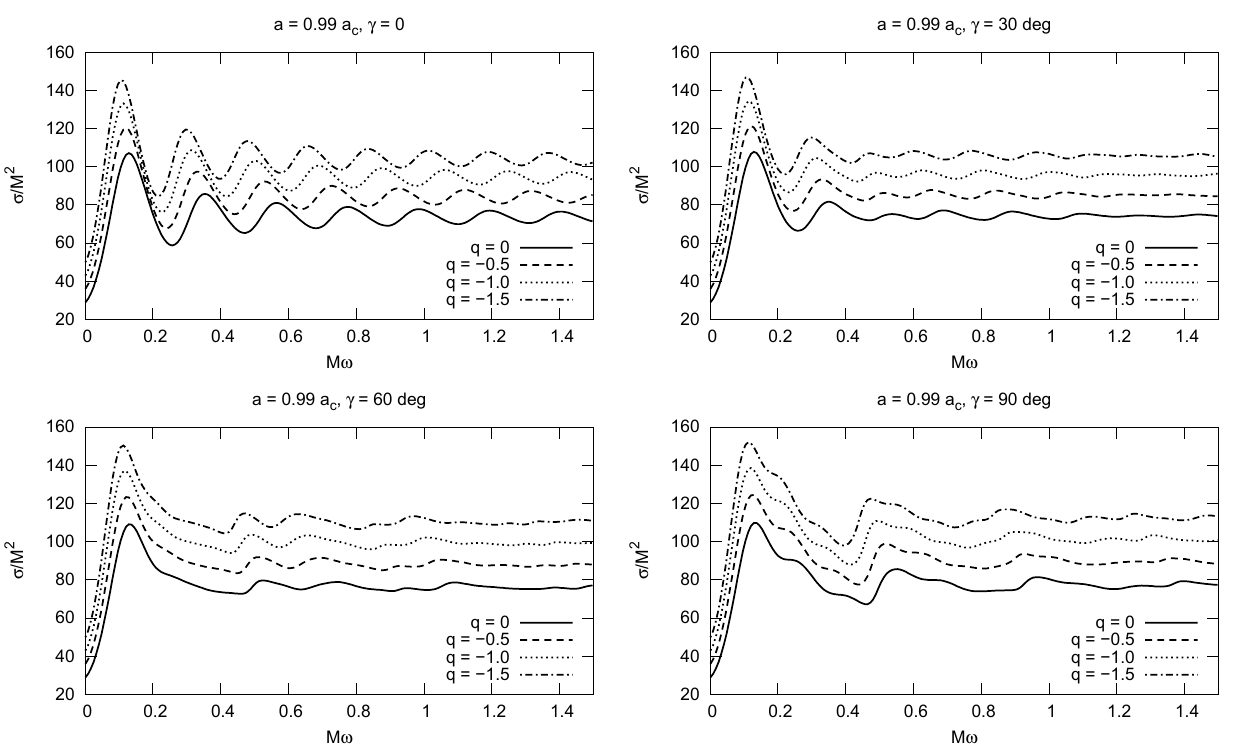}
	\caption{Total absorption cross sections of rotating black holes with tidal charges for massless scalar waves. Here we consider near-extreme black holes ($a = 0.99\,a_c$) with $q = 0, -0.5, -1.0, -1.5$ and four different angles of incidence: $\gamma = 0$ (top-left), $\gamma = 30\text{ deg}$ (top-right), $\gamma = 60\text{ deg}$ (bottom-left), and $\gamma = 90\text{ deg}$ (bottom-right).}
	\label{fig:acs_near_ext}
\end{figure*}

In Fig.~\ref{fig:acs_near_ext} we show the total absorption cross section of near-extreme rotating black holes with $q = 0, -0.5, -1.0, -1.5$ for massless scalar waves impinging upon the black hole from $\gamma = 0, 30\text{ deg}, 60\text{ deg}, 90\text{ deg}$. Similarly to what happens with the capture cross section, the increase of the negative tidal charge intensity results in an increase of the total absorption cross section for all frequencies. The absorption spectra presents an interesting similarity for near-extreme black holes with different tidal charges if we consider the same angle of incidence. The parameters chosen for the black holes and incidence directions in Fig.~\ref{fig:acs_near_ext} are the same of the ones for the shadows shown in Fig.~\ref{fig:shad_comp}. Comparing both figures we conclude that there is a relation between the shadow circularity and the regularity of oscillation which the absorption cross section presents.

\begin{figure}[!htb]
	\centering
	\includegraphics[width=0.49\textwidth]{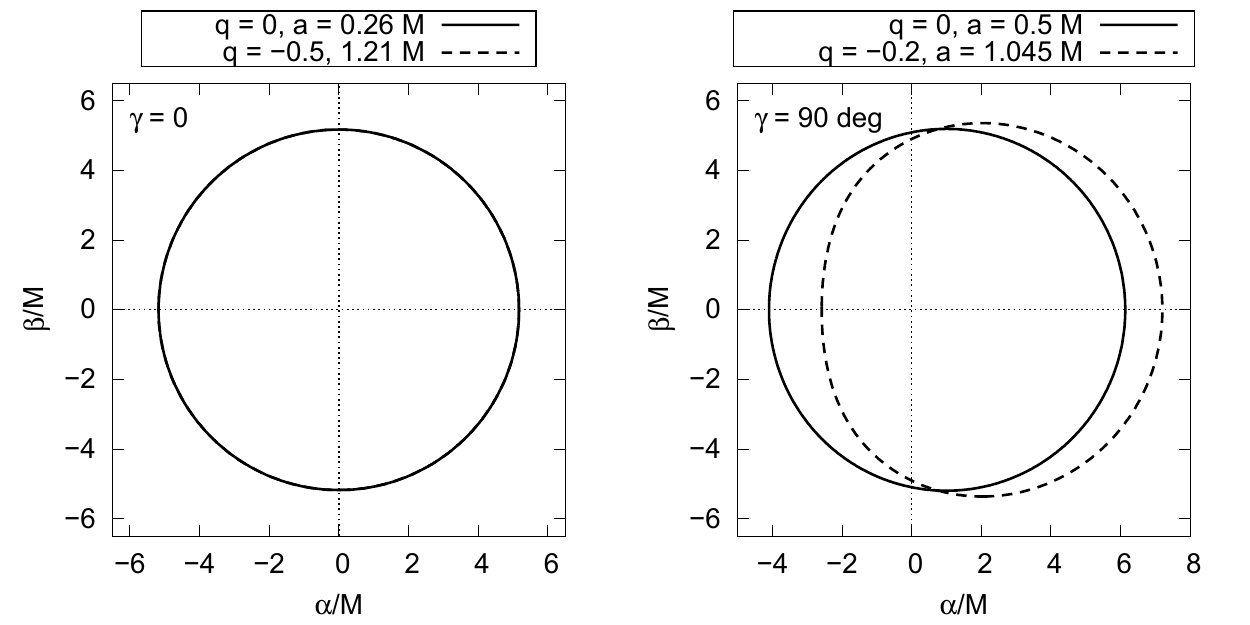}
	\caption{Examples of equal-size shadows for black holes with different configurations. \emph{Left:} shadows as seen by a very distant observer at $\gamma = 0$ considering Kerr black holes with $a = 0.26 M$, and near-extreme rotating black holes with tidal charge $q = -0.5$. \emph{Right:} shadows of Kerr black holes with $a = 0.5 M$, and rotating black holes with tidal charge $q = -0.2$, $a = 1.045 M$ as seen by a very distant observer at the equatorial plane.}
	\label{fig:equiv_shad}
\end{figure}

\begin{figure}[!htb]
	\centering
	\includegraphics[width=0.49\textwidth]{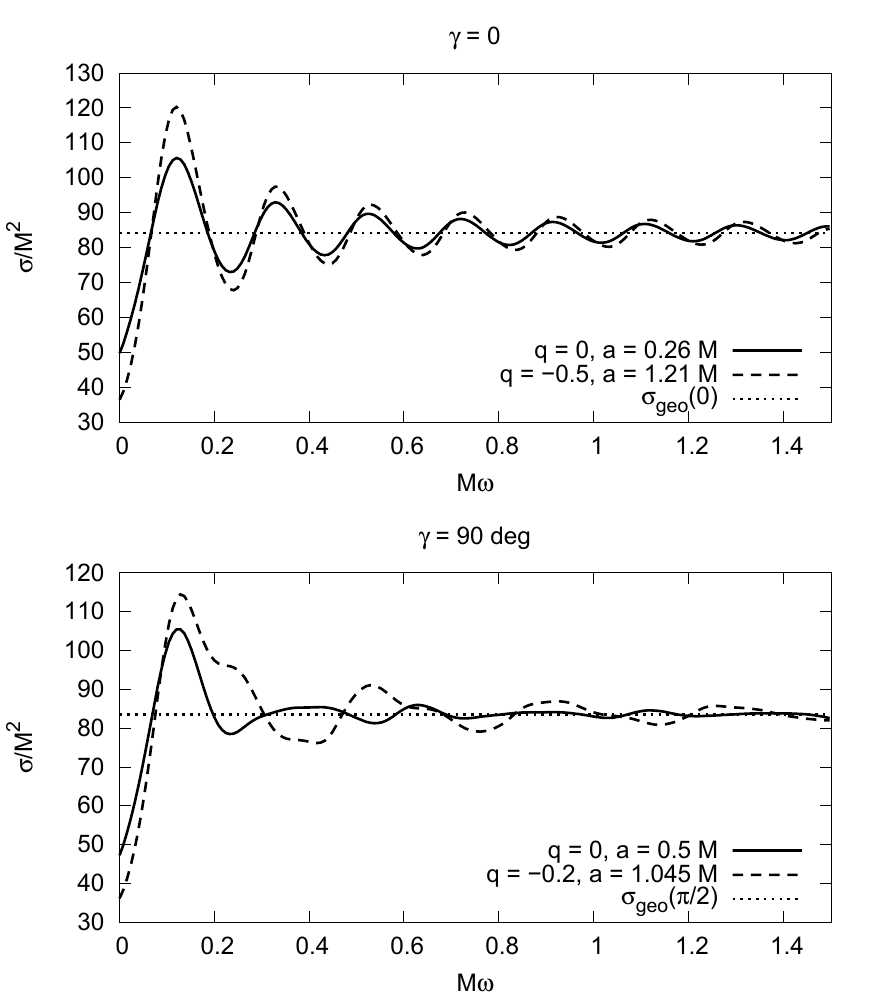}
	\caption{Absorption cross sections for black holes which possess capture cross sections of same size. \emph{Top:} the case of rotating black holes with tidal charges $q = 0, -0.5$ and $a/M = 0.26, 1.21$, respectively, considering axial incidence. \emph{Bottom:} The case of rotating black holes with tidal charges $q = 0, -0.2$ and $a/M = 0.5, 1.045$, respectively, considering an incidence along $\gamma = 90$ deg.}
	\label{fig:equiv_acs}
\end{figure}

The increases of $-q$ and $a/M$ have opposite effects at the absorption cross section average value; while the increase of $-q$ with $a/M$ fixed is followed by an increase of the absorption cross section average (cf. Fig.~\ref{fig:a0.99acs}), the increase of $a/M$ when the value of $q$ is unaltered results in a decrease of this average (cf. Fig.~\ref{fig:a_acs}). This indicates the possibility of black holes with different configurations possessing capture cross sections of same size. If this is the case, it is interesting to analyze the level of similarity that the absorption cross sections of such black holes present. Figure~\ref{fig:equiv_shad} shows an example of shadows of same size for black holes with different configurations. For $\gamma = 0$, we have chosen the cases $q = 0, -0.5$ with $a/M = 0.26, 1.21$, respectively, while for $\gamma = 90$ deg, $q = 0, -0.2$ with $a/M = 0.5, 1.045$, respectively. For $\gamma = 0$, the black holes cannot be distinguished by the size or shape of their shadow, while for $\gamma = 90$ deg, although both shadows have the same area, their shape is evidently distinct. The absorption cross sections for the same sets ($q,a,\gamma$) are presented in Fig.~\ref{fig:equiv_acs}. For $\gamma = 0$, despite very similar, the absorption cross sections are different, mainly for small frequencies; black holes with tidal charge present absorption cross sections with higher oscillations and out of phase if compared with the absorption cross section of Kerr black holes. For $\gamma = 90$ deg, similarly to what happens with the shadows, the difference between the absorption cross sections becomes more evident. The black hole with less circular shadow presents also the absorption cross section with less regular oscillatory behavior, as we have observed previously.

\section{Final remarks}
\label{sec:conc}

We have presented the analysis of the absorption cross section of rotating black holes with negative tidal charges by considering the scattering of monochromatic plane massless scalar waves. Although we have focused our analysis in the cases of near-extreme black holes ($a = 0.99 a_c$) with specific values of tidal charge for selected directions of incidence, it can be directly generalized to any other allowed values of $a$, $q$, and $\gamma$. In most of our analysis, we have compared the absorption cross sections of rotating black holes with tidal charges with the Kerr case which has been extensively analyzed in Ref.~\cite{Macedo2013prd88_064033}.

The influence of the tidal charge is observed specifically in the average size of the absorption cross section, which increases if the negative tidal charge intensity increases considering black holes with fixed values of $a/M$ (Fig.~\ref{fig:a0.99acs}), or near-extreme black holes (Fig.~\ref{fig:acs_near_ext}). The absorption spectrum for off-axis incidence of near-extreme black holes with different tidal charges can be surprisingly similar in shape, although with different sizes (Fig.~\ref{fig:acs_near_ext}), showing that the form of the absorption spectrum is strongly related to the value of $a/a_c$.

The fact that the increase of the negative tidal charge intensity results in an increase of the absorption cross section for near-extreme black holes has also been noted in the partial absorption. This also applies to negative absorption due to superradiance. However, differently from the amplification factors, of which maximum value can increase with the increase of $l = m$ for rotating black holes with tidal charges $-q \gtrsim 2$~\cite{deOliveira2020lzp}, the partial absorption cross section gets highly suppressed in this sense. This reveals the contrast between the geometric characteristic of the absorption cross section and the fact that greybody/amplification factors measure rates.

There is a relation between the circularity of the shadow and the regularity of the absorption cross section oscillations around the respective high-frequency limit, which is the value of the capture cross section. As the shadow loses circularity, the corresponding absorption cross section becomes more irregular (Figs.~\ref{fig:equiv_shad} and~\ref{fig:equiv_acs}). For an observer at $\gamma = 0$, the shadows cast by rotating black holes with tidal charge will be circular, what indicates the possibility of different black holes presenting identical shadows. Even in such cases, the absorption cross section will present differently for black holes with different configurations mainly in the small-frequency regime.

Results were compared with the low-frequency approximation~\cite{Higuchi2001cqg18_L139} (Fig.~\ref{fig:pacs0}), and with the respective capture cross sections area (Fig.~\ref{fig:a0.99acs}) resulting in good agreements. We have used this agreement together with convergence tests to check the consistency of our numeric evaluations.

\section*{Acknowledgments}

The author would like to thank Conselho Nacional de Desenvolvimento Científico e Tecnológico (CNPq) for partial financial support via the grants 304679/2018-6 and 427532/2018-3.

%\bibliographystyle{unsrtnat}
%\bibliography{references.bib}

\end{document}